\documentclass[12pt]{article}
\usepackage{graphicx}
\usepackage{url} % not crucial - just used below for the URL 

\usepackage[utf8]{inputenc}
\usepackage{amsfonts}
\usepackage{amsmath}
\usepackage{amssymb}
\usepackage[english]{babel}
\usepackage{bbm}
\usepackage{csquotes}
\usepackage{relsize}

\usepackage{natbib} %comment out if you do not have the package

\allowdisplaybreaks

\newcommand{\blind}{0}

\begin{document}

\def\spacingset#1{\renewcommand{\baselinestretch}%
{#1}\small\normalsize} \spacingset{1}

%%%%%%%%%%%%%%%%%%%%%%%%%%%%%%%%%%%%%%%%%%%%%%%%%%%%%%%%%%%%%%%%%%%%%%%%%%%%%%

\if0\blind
{
  \title{\bf Adaptively Sampling via Regional Variance-Based Sensitivities}
  \author{
    Brian\ W.\ Bush \\
    National Renewable Energy Laboratory, \\ \\
    Joanne Wendelberger \\
    Los Alamos National Laboratory, \\ \\
    Rebecca Hanes \\
    National Renewable Energy Laboratory
  }
  \maketitle
} \fi

\if1\blind
{
  \bigskip
  \bigskip
  \bigskip
  \begin{center}
    {\LARGE\bf Adaptively Sampling via Regional Variance-Based Sensitivities}
\end{center}
  \medskip
} \fi

%%%% FIXME: Remove comment before submitting.
%\spacingset{2} % DON'T change the spacing!

\bigskip
\begin{abstract}
Inspired by the well-established variance-based methods for global sensitivity analysis, we develop a local total sensitivity index that decomposes the global total sensitivity conditions by independent variables' values. We employ this local sensitivity index in a new method of experimental design that sequentially and adaptively samples the domain of a multivariate function according to local contributions to the global variance. The method is demonstrated on a nonlinear illustrative example that has a three-dimensional domain and a three-dimensional codomain, but also on a complex, high-dimensional simulation for assessing the industrial viability of the production of bioproducts from biomass.
%\todo{\small 100 or fewer words.}
\end{abstract}

\noindent%
{\it Keywords: Computer Experiments; Experimental Design; Multivariate Analysis; Sampling.}
%\todo{\small 3 to 6 keywords, that do not appear in the title}
\vfill

\newpage

\section{Introduction}
\label{sec:intro}

Industrially relevant computer experiments employing complex simulations, which typically involve dozens or hundreds of independent and dependent variables, pose daunting analytic challenges for developing defensible and robust insights about sensitivity, uncertainty, and system behavior. The high dimensionality of the input spaces necessitates economical exploration of input-parameter ranges and the similarly large output dimensions, often in the form of multivariate time series, stress visual and statistical analyses. Well-established techniques for global sensitivity analysis, such as elementary-effects screening and variance-based first-order and total sensitivity indices \citep{saltelli_global_2008}, provide a solid foundation for exploring and characterizing these complex simulations, while local or regional methods such as Monte-Carlo filtering \citep{borgonovo_sensitivity_2016} allow targeted discrimination of sensitivities, but neither approach maps the landscape of contributions to sensitive dependence of output variabilities upon input values. Understanding that landscape of sensitivity, however, may motivate deeper exploration and an iterative sequence of computer experiments to tease out significant details of the nonlinear responses in a simulation.

Indeed, a not uncommon practice in computer experimentation on complex models or simulations relies on an initial screening of sensitivities over a generously large set of variables, typically chosen according to the judgment of model developers and domain experts:\ the result of the broad, but computationally efficient, screening identifies a set of candidate input parameters for which a more rigorous and expensive global sensitivity analysis is warranted. Initial screening may use one-at-a-time variation of input parameters, while the subsequent global analysis typically relies on a uniformly-sampling space-filling design (e.g., fractional factorial, Latin hypercube, orthogonal array, quasi-random sequence, or a hybrid of these).
%\todo{\small Do we need to cite all of the bread-and-butter techniques mentioned in this paragraph?}
Based on the results of the global sensitivity analysis and resources permitting, an analyst may execute subsequent sensitivity analyses that (1) narrow or broaden the range of variation in the independent variables, (2) add or remove candidate inputs from the set varied in the experiments, (3) add or remove output variables for evaluating sensitivities, or (4) increase sample size. Sample size is a particularly important issue because of the inverse relationship between number of input parameters varied in a sensitivity analysis and the tightness of the confidence intervals in the sensitivity indices (especially for total sensitivities or for higher than first-order sensitivities). Thus, subsequent iterations of a global sensitivity analysis may reduce the number of inputs considered, in order to tighten confidence intervals for the inconclusive results of previous iterations, and increase sample size. Early iterations might also correct overly broad or narrow ranges of input variation or discard outputs that do not exhibit interesting variation, as when a modeler or analyst belatedly realizes that an output variable exhibits a trivial dependence on one or more inputs or trivially inconsequential variation. Furthermore, exploratory analysis and visualization of the raw simulation output, conditioned on the inputs, may reveal interesting patterns and dependencies---not apparent in the global sensitivity indices---of the output upon local regions of the input space:\ this may lead to focused ``side analyses'' that probe local regions or test specific hypotheses. If computational resources were unlimited, then none of this tedious and labor-intensive iteration would be necessary.

Our work here is motivated by the desire to automate much of the practical and predictably iterative process of model exploration and sensitivity analysis described in the previous paragraph, thus freeing the modeler and analyst to work at a higher level of abstraction in developing insights and testing hypotheses. In particular, we would like to semi-autonomously focus computational resources on the most potentially interesting regions of the input space using estimates of how much those regions contribute to global sensitivity indices. We envision a computational workflow where (1) an initial uniformly space-filling experimental design adaptively evolves to focus on regions of higher variance while not altogether ignoring the lower-variance regions, (2) previously computed simulations are seamlessly incorporated into the adaptive design, (3) users may at any point tune the intensity of focus on the higher-variance regions, and (4) users have the option to intervene to adjust the boundaries of the input space or the density of sampling.
%\todo{Should we add a workflow diagram illustrating this?}
To this end, in this paper we develop a local decomposition of global variance-based sensitivity indices, formulated in a manner that enables iterative updating of the non-uniform space-filling sampling of the simulation output so that it focuses on higher-variance regions. Additionally, the sampling method is amenable to user interventions to tailor sampling to their evolving interests.

\section{Background}
\label{sec:back}

Saltelli et al.\  (\citeyear{saltelli_sensitivity_2004,saltelli_global_2008}) provide thorough and practical overviews of global variance-based sensitivity analysis methods, while their later work \citep{saltelli_variance_2010} summarizes recommendations for designing experiments for and computing first-order and total sensitivity indices. Such recommendations arise from foundational work by a variety of researchers who investigated optimal experimental design \citep{jansen_analysis_1999,saltelli_variance_2010}, alternative computational recipes \citep{saltelli_variance_2010}, and efficient organization and minimization of model evaluations \citep{saltelli_making_2002,piano_new_2019,kucherenko_estimation_2012}.

Borgonovo and Plischke \citet{borgonovo_sensitivity_2016} broadly catalog the spectrum of sensitivity-analysis techniques, including not only global variance-based methods, but also local or regional non-variance methods such as Monte Carlo filtering \citep{wu_application_2017}. Domain-specific reviews and assessments of sensitivity analysis are provided by \citet{wagener_what_2019} for earth-system modeling, \citet{tian_review_2013} for building-energy analysis, \citet{norton_introduction_2015} or \citet{pianosi_sensitivity_2016} for environmental-simulation models, and \citet{jadun_application_2017} or \citet{inman_application_2018} for bioenergy supply-chain simulation.

Local and regional specializations of sensitivity analysis are discussed in \citet{wei_regional_2014}, \citet{wu_application_2017}, \citet{spear_parameter_1994}, and \citet{rose_parameter_1991}. \citet{gamboa_sensitivity_2013} generalize sensitivity indices to multivariate output. \citet{lu_non-uniform_2020} propose non-uniform space-filling (NUSF) experimental designs that control the density of sampling according to experimental objectives, while \citet{bowman_weighted_2013} define weighted space-filling (WSF) ones using a distance metric for the design space.

% Discuss Myers et al. \cite{myers_partitioning_2016}?

\section{Method}
\label{sec:meth}

We consider multivariate functions $\mathbf{Y} \left( \mathbf{X} \right)$ with $m$ independent variables $X_i$ and $n$ dependent variables $Y_j$. (Table~\ref{tab:notation} summarizes the notation used in this paper.) These functions may represent the input-output relationship of computer models or simulations (either deterministic or stochastic) or physical experiments.

We approach the problem of developing local versions of variance-based global sensitivity indices by expressing the standard global indices \citep{saltelli_variance_2010} in forms amenable to local decomposition. A local total sensitivity separately partitions the global total sensitivity along each independent variable. We then adopt experimental designs that are suitable for iterative extension. The iterative experimental design samples the space of independent variables proportionally to the local variance indices observed so far in the experiment.

\begin{table}
    \centering
    \begin{tabular}{ll}
        Symbol & Definition \\
        \hline
        $i \in \{1 \ldots m\}$ & independent-variable index and dimension \\
        $j \in \{1 \ldots n\}$ & dependent-variable index and dimension \\
        $k \in \{1 \ldots N\}$ & observation index and dimension \\
        $\mathbb{E}_Z$ & expectation over random variable(s) $Z \in \{ X_i, \mathbf{X}_{\sim i} \}$ \\
        $\mathbb{V}_Z$ & variance over random variable(s) $Z \in \{ X_i, \mathbf{X}_{\sim i} \}$ \\
        $X_i \in [0, 1]$ & independent variable for dimension $i$ \\
        $\mathbf{X}_{\sim i}$ & independent variables except for dimension $i$ \\
        $Y_j$ & dependent variable for dimension $j$ \\
        $\mathbf{A}$, $\mathbf{B}$ & $N \times k$ design matrices \\
        $\mathbf{A}_{\mathbf{B}_i}$ & design matrix $\mathbf{A}$ with its $i$th column replaced \\ & by the $i$th column of $\mathbf{B}$ \\
        $x_{\mathbf{Z}_i}^{(k)} \in [0, 1]$ & $i$th independent variable for the $k$th observation \\ & in the design matrix $\mathbf{Z} \in \{ \mathbf{A}, \mathbf{B} \}$ \\
        $y_{\mathbf{Z},j}^{(k)}$ & $j$th dependent variable for the $k$th observation \\ & using the design matrix $\mathbf{Z} \in \{ \mathbf{A}, \mathbf{B}, \mathbf{A}_{\mathbf{B}_i} \}$ \\
        $\mathbbm{1}_c$ & indicator function: $1$ if the condition $c$ holds, \\ & but $0$ otherwise \\
        $u \mathrel{\hat=} v$ & $v$ is an estimate for $u$ \\
    \end{tabular}
    \caption{Notation used in this paper, inspired by and generally consistent with \citet{saltelli_variance_2010}.}
    \label{tab:notation}
\end{table}

\subsection{Global Sensitivity}
\label{sec:glob}

We start from standard definitions \citep{saltelli_variance_2010} that use conditional expectations and variances to express the first-order and total sensitivity indices, respectively:
\begin{align}
    S_{i,j} & = \frac{\mathbb{V}_{X_i} \left[ \mathbb{E}_{\mathbf{X}_{\sim i}} \left[ Y_j \middle| X_i \right] \right]}{\mathbb{V}_\mathbf{X} \left[ Y_j \right]} \label{eq:satellis}
    \\
    T_{i,j} & = \frac{\mathbb{E}_{\mathbf{X}_{\sim i}} \left[ \mathbb{V}_{X_i} \left[ Y_j \middle| \mathbf{X}_{\sim i} \right] \right]}{\mathbb{V}_\mathbf{X} \left[ Y_j \right]} \label{eq:saltellit}
\end{align}
Loosely speaking, the first-order index $S_{i,j}$ quantifies the influence of $X_i$ solely upon the variance of $Y_j$ whereas the total index $T_{i,j}$ quantifies the influence of $X_i$ in combination with other independent variables upon the variance of $Y_j$. These two indices arise from decomposing the variance $Y_j$ and they obey a variety of summation relationships and bounds---see \citet{saltelli_sensitivity_2004} for detailed discussion.

Practically estimating Eqs.~(\ref{eq:satellis}) and~(\ref{eq:saltellit}) may be problematic because of computational complexity, finite sample size, and numerical issues. Careful design of experiments alleviates some of these difficulties. An efficient design employs two $N \times k$ design matrices, $\mathbf{A}$ and $\mathbf{B}$, that are hybridized into matrices $\mathbf{A}_{\mathbf{B}_i}$ that are constructed by replacing the $i$th column of $\mathbf{A}$ with the $i$th column of $\mathbf{B}$. The number of rows $N$ determines the size of the experiment, which requires evaluating $\mathbf{Y}\left(\mathbf{X}\right)$ for each row of $\mathbf{A}$, $\mathbf{B}$, and $\mathbf{A}_{\mathbf{B}_i}$: thus $N (k + 2)$ observations must be made. This column-swapping construction facilitates efficient evaluation of the conditional expectations and variances by controlling which independent variables are varied and by organizing the computation.

The best-practice recommendations of \citet{saltelli_variance_2010} are to use the formulae in \citet{jansen_analysis_1999} when computing the variances of conditional expectations and the expectations of conditional variances:
\begin{align}
    \mathbb{V}_\mathbf{X} \left[ Y_j \right] - \mathbb{V}_{X_i} \left[ \mathbb{E}_{\mathbf{X}_{\sim i}} \left[ Y_j \middle| X_i \right] \right] & \mathrel{\hat=} \frac{1}{2 N} \sum_{k=1}^N \left| y_{\mathbf{B},j}^{(k)} - y_{\mathbf{A}_{\mathbf{B}_i},j}^{(k)} \right|^2 \label{eq:jansens}
    \\
    \mathbb{E}_{\mathbf{X}_{\sim i}} \left[ \mathbb{V}_{X_i} \left[ Y_j \middle| \mathbf{X}_{\sim i} \right] \right] & \mathrel{\hat=} \frac{1}{2 N} \sum_{k=1}^N \left| y_{\mathbf{A},j}^{(k)} - y_{\mathbf{A}_{\mathbf{B}_i},j}^{(k)} \right|^2 \label{eq:jansent}
\end{align}
Here the $y_{\mathbf{A},j}^{(k)}$, $y_{\mathbf{B},j}^{(k)}$, and $y_{\mathbf{A}_{\mathbf{B}_i},j}^{(k)}$ are the observation of $Y_j$ for the $k$th row of the design matrix $\mathbf{A}$, $\mathbf{B}$, and $\mathbf{A}_{\mathbf{B}_i}$, respectively. We supplement these formulae with the analogous formula for computing the overall variance:
\begin{equation}
    \mathbb{V}_\mathbf{X} \left[ Y_j \right] \mathrel{\hat=} \frac{1}{2 N} \sum_{k=1}^N \left| y_{\mathbf{A},j}^{(k)} - y_{\mathbf{B},j}^{(k)} \right|^2 \label{eq:jansenv}
\end{equation}

Combining Eqs.~(\ref{eq:jansens}) through (\ref{eq:jansenv}), we adopt the following estimators for the first-order and total global sensitivities and the variance:
\begin{align}
    \hat{S}_{i,j} = 1 - & \frac{\sum_{k=1}^N \left| y_{\mathbf{B},j}^{(k)} - y_{\mathbf{A}_{\mathbf{B}_i},j}^{(k)} \right|^2}{\sum_{k=1}^N \left| y_{\mathbf{A},j}^{(k)} - y_{\mathbf{B},j}^{(k)} \right|^2} \label{eq:ours}
    \\
    \hat{T}_{i,j} = & \frac{\sum_{k=1}^N \left| y_{\mathbf{A},j}^{(k)} - y_{\mathbf{A}_{\mathbf{B}_i},j}^{(k)} \right|^2}{\sum_{k=1}^N \left| y_{\mathbf{A},j}^{(k)} - y_{\mathbf{B},j}^{(k)} \right|^2} \label{eq:ourt}
    \\
    \hat{V}_{j} = & \frac{1}{2 N} \sum_{k=1}^N \left| y_{\mathbf{A},j}^{(k)} - y_{\mathbf{B},j}^{(k)} \right|^2 \label{eq:ourv}
\end{align}

\subsection{Experimental Design}
\label{sec:design}

Although any of a number of design methods could be used for creating the $\mathbf{A}$ and $\mathbf{B}$ matrices, quasi-random sequences are well suited for computing the sensitivity indices and \citet{saltelli_variance_2010} recommends using Sobol' sequences because of their low discrepancy properties. In this procedure, the $\mathbf{A}$ and $\mathbf{B}$ matrices are placed side-by-side to form an $N \times 2 k$ matrix where the $N$ rows are consecutive points in the $2 k$-dimensional Sobol' sequence. An experiment can very simply be enlarged just by appending additional rows of the Sobol' sequence, so this design procedure is well suited for iteratively extending a sensitivity analysis.

Without loss of generality, we require that the domain of each independent variable be $[0, 1]$. This is convenient because the range of the Sobol' sequence is also $[0, 1]$ in each dimension.

\subsection{Local Sensitivity}
\label{sec:local}

In developing a local version of variance-based sensitivity, one would like to attribute the variance in output to a specific value or interval of the independent variables. Equation~(\ref{eq:ourt}), for example, involves evaluating $Y_j$ at the points $x_{\mathbf{A}_i}^{(k)}$ and $x_{\mathbf{B}_i}^{(k)}$ to compute the contribution $\left| y_{\mathbf{A},j}^{(k)} - y_{\mathbf{A}_{\mathbf{B}_i},j}^{(k)} \right| ^2$ to the global total sensitivity. For lack of a more specific way to assign this contribution to a particular value of the independent variable, we can simply spread this contribution uniformly between $x_{\mathbf{A}_i}^{(k)}$ and $x_{\mathbf{B}_i}^{(k)}$.
%\todo{Should we add a diagram showing this?}
Thus we define $t_{i,j}^{(\alpha,\epsilon)}(x_i)$ to be the local sensitivity of the output $y_j$ upon the input $x_i$:
\begin{equation}
    t_{i,j}^{(\alpha,\epsilon)}(x) = \mathlarger{\mathlarger{\sum}}_{k=1}^N \left| y_{\mathbf{A},j}^{(k)} - y_{\mathbf{A}_{\mathbf{B}_i},j}^{(k)} \right| ^ \alpha  \cdot \frac{\mathbbm{1}_{\min \left\{ x_{\mathbf{A}_i}^{(k)} , x_{\mathbf{B}_i}^{(k)} \right\} - \frac{\epsilon}{2} \leq x \leq \max \left\{ x_{\mathbf{A}_i}^{(k)} , x_{\mathbf{B}_i}^{(k)} \right\} + \frac{\epsilon}{2}}}{\left| x_{\mathbf{A}_i}^{(k)} - x_{\mathbf{B}_i}^{(k)} + \epsilon \right|} \label{eq:tx}
\end{equation}
Here we have introduced $\epsilon > 0$ to guard against division by zero, which could occur if the two $x_i$ coincide, and we have replaced the exponent $2$ by the general parameter $\alpha$. The parameter $\alpha$ amplifies the difference between the dependent variables, with $\alpha = 2$ corresponding to computing the variance. The parameter $\epsilon$ slightly widens the interval over which the contribution to the variance is spread. It is also useful to define a cumulative version of the local sensitivity:
\begin{equation}
    T_{i,j}^{(\alpha,\epsilon)}(x) = \frac{1}{2 N \hat{V}_j} \int_{-\infty}^x d x^\prime \; t_{i,j}^{(\alpha,\epsilon)}(x^\prime) \label{eq:cumt}
\end{equation}
Note that we can recover the global total sensitivity index, Equation (\ref{eq:ourt}), from the limit of Equation (\ref{eq:cumt}):
\begin{equation*}
    \hat{T}_{i,j} = \lim_{x \rightarrow \infty} T_{i,j}^{(2,\epsilon)}(x)
\end{equation*}

In order to sample an independent variable proportionally to $t_{i,j}^{(\alpha,\epsilon)}(x_i)$, we need a normalized version of it, which we call the sensitivity density:
\begin{equation}
    \tau_{i,j}^{(\alpha,\epsilon)}(x) = \left. \frac{t_{i,j}^{(\alpha,\epsilon)}(x)}{t_{i,j}^{(0,\epsilon)}(x)} \middle/ \int_{-\infty}^{\infty} d x^\prime \; \frac{t_{i,j}^{(\alpha,\epsilon)}(x^\prime)}{t_{i,j}^{(0,\epsilon)}(x^\prime)} \right. \label{eq:taux}
\end{equation}
The factor $t_{i,j}^{(0,\epsilon)}(x)$ in the denominators removes the non-uniformity in the sampling of $x_i$. Finally, we can further summarize the local sensitivity by averaging it over the $n$ dependent variables; we call this the average sensitivity density:
\begin{equation}
    \overline{\tau}_i^{(\alpha,\epsilon)}(x) = \frac{1}{n} \sum_{j=1}^n \tau_{i,j}^{(\alpha,\epsilon)}(x) \label{eq:taubar}
\end{equation}
It also has a cumulative version:
\begin{equation}
    \overline{T}_i^{(\alpha,\epsilon)}(x) = \int_{-\infty}^x d x^\prime \; \overline{\tau}_i^{(\alpha,\epsilon)}(x^\prime) \label{eq:tbar}
\end{equation}

Unfortunately, the expression of the first-order sensitivity in Eq.~(\ref{eq:ours}) as the difference of two terms precludes its analogous decomposition into a local first-order sensitivity index. One could create a local first-order \textit{insensitivity} index based on $1 - \hat{S}_{i,j}$, but the density of insensitivity would not be practical for iteratively adapting experimental designs.

\subsection{Adaptively Iterating Sensitivity Analysis}
\label{sec:adapt}

We are now in a position to specify an algorithm for iteratively adapting the experimental design for a sensitivity analysis as that analysis proceeds. Let $M$ be the batch size for the iterations.
\begin{enumerate}
    \item Evaluate $y_{\mathbf{A},j}^{(k)}$, $y_{\mathbf{B},j}^{(k)}$, and $y_{\mathbf{A}_{\mathbf{B}_i},j}^{(k)}$ for the first $M$ rows of the design matrices $\mathbf{A}$, $\mathbf{B}$, and $\mathbf{A}_{\mathbf{B}_i}$.
    \item Compute the cumulative average sensitivity densities $\overline{T}_i^{(\alpha,\epsilon)}(x)$ for each independent variable.
    \item Append $M$ rows to the design matrices by taking the next $M$ points in the Sobol' sequence, but transform those points according to the average sensitivity density. Namely, the Sobol' point $s_\ell$ is mapped to ${\overline{T}_i^{(\alpha,\epsilon)}}^{-1}(s_\ell)$, where $i = 1 + (\ell - 1 \mod m)$, since $\ell \in \{ 1 \ldots 2 m \}$ because the $\mathbf{A}$ and $\mathbf{B}$ matrices are placed side-by-side in the Sobol' space.
    \item Evaluate $y_{\mathbf{A},j}^{(k)}$, $y_{\mathbf{B},j}^{(k)}$, and $y_{\mathbf{A}_{\mathbf{B}_i},j}^{(k)}$ for these $M$ new rows of the design matrix.
    \item Proceed to step \#2 above.
\end{enumerate}
For computational efficiency, the $\hat{V}_j$ and $T_{i,j}^{(\alpha,\epsilon)}(x)$ can be updated during each iteration instead of being computed from scratch. The batch size can be $M = 1$ if an online (i.e., serial piece-by-piece) algorithm is desired. The parameter $\epsilon$ should be set to a small positive value. The parameter $\alpha$ allows one to make the sampling completely non-adapative ($\alpha = 0$) or highly adaptive (large positive $\alpha$): setting $\alpha$ too high may result in missing sensitivities in the input space if the batch size is too small, since the algorithm may home in on the first observed region of sensitivity, thus missing other areas of high sensitivity. After the prominent regions of sensitivity had been mapped with $\alpha > 0$, one could even use a negative $\alpha$ in subsequent iterations to search the nominally insensitive regions of the input space for hidden sensitivities that were not previously detected due to insufficiently dense sampling.

\section{Results}
\label{sec:result}

We evaluate the adaptive iterative approach to sensitivity analysis first using a synthetic example model created for demonstration purposes, which has a manageably small number of dimensions, and then using a complex simulation developed for industry analyses.

\subsection{Synthetic Example Model}
\label{sec:toy}

Appendix~\ref{sec:toyapp} describes a class of synthetic example models that have strong nonlinearities, configurable dimensionality, and specifiable discontinuities in the dependent variables. For convenience in exploring, visualizing, and plotting results, we select three input dimensions ($m = 3$) and three output dimensions ($n = 3$). To capture different types of behavior encountered in investigations of computer models used in industrial applications, the model has a zeroth-order discontinuity in the output at $x_1 \doteq 0.59$, a first-order discontinuity in the output at $x_2 \doteq 0.95$, and a second-order discontinuity in the output at $x_3 \doteq 0.10$.

\begin{figure}
    \centering
    \includegraphics[width=1.00\linewidth]{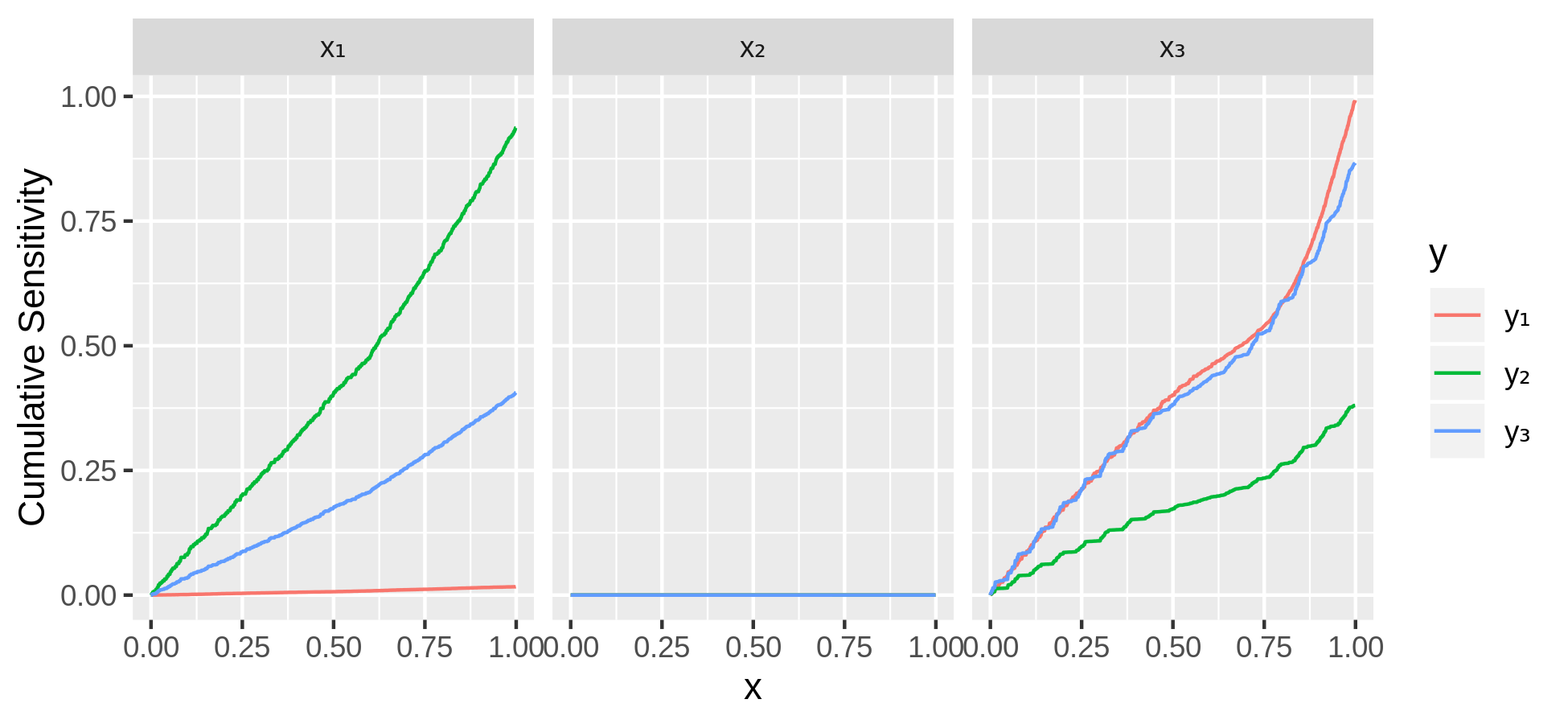}
    \caption{Cumulative local sensitivity $T_{i,j}^{(2,\epsilon)}(x)$ for the synthetic example model, with $\epsilon = 10^{-4}$ and $N = M = 1000$.}
    \label{fig:toy-tx}
\end{figure}

\begin{figure}
    \centering
    \includegraphics[width=1.00\linewidth]{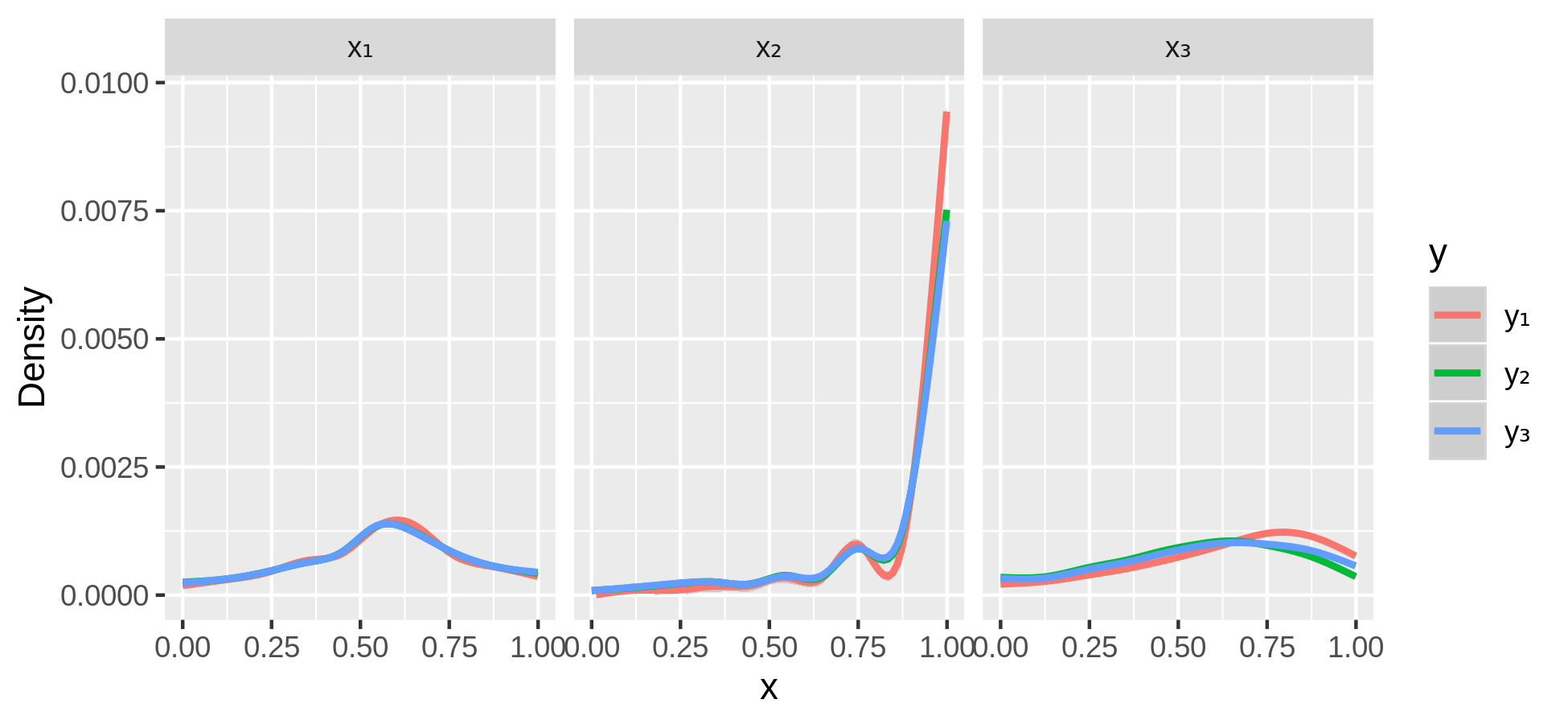}
    \caption{Sensitivity density $\tau_{i,j}^{(2,\epsilon)}(x)$ for the synthetic example model, with $\epsilon = 10^{-4}$ and $N = M = 1000$. (Note that the density is smoothed using a general additive model.)}
    \label{fig:toy-tau}
\end{figure}

\begin{figure}
    \centering
    \includegraphics[width=1.00\linewidth]{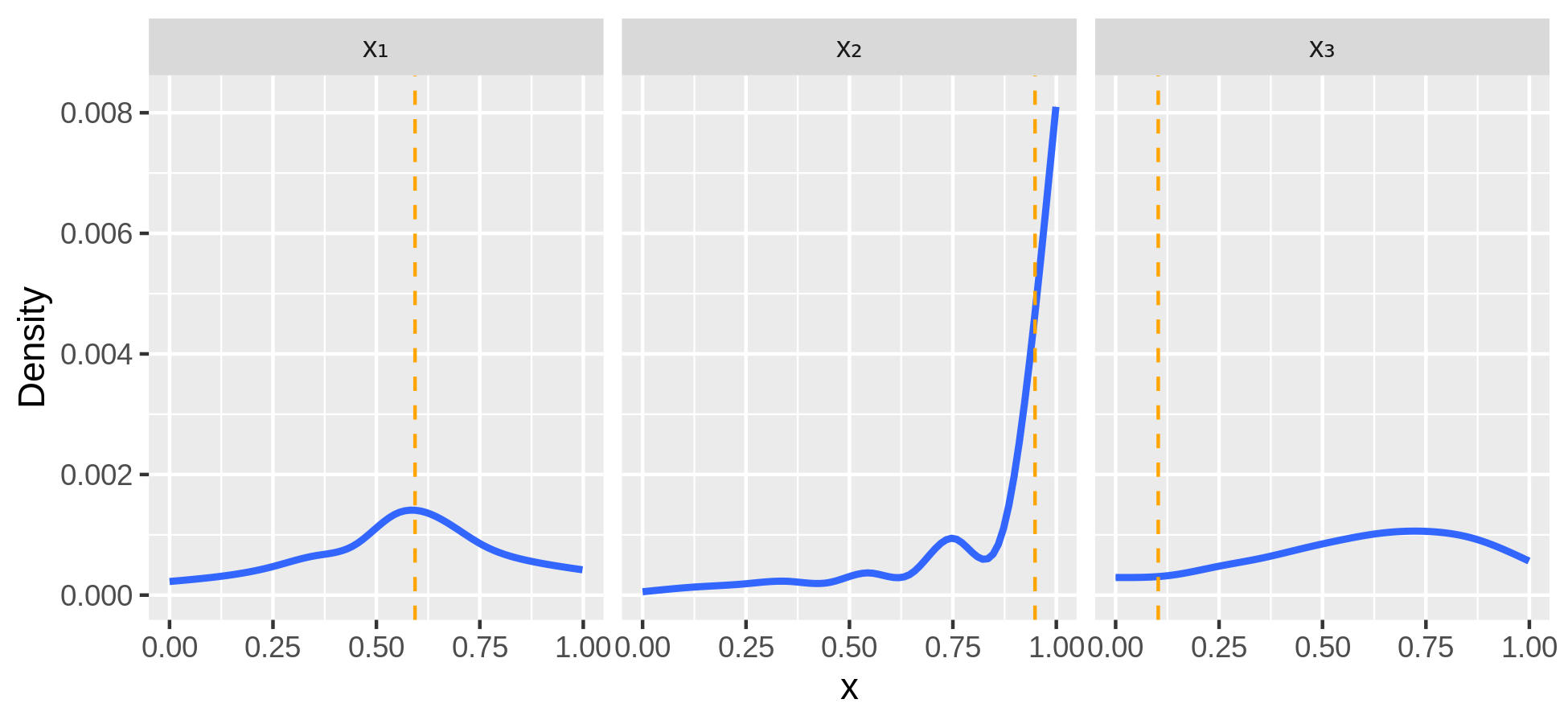}
    \caption{Averaged sensitivity density $\overline{\tau}_i^{(2,\epsilon)}(x)$ for the synthetic example model, with $\epsilon = 10^{-4}$ and $N = M = 1000$. The three dashed orange lines mark locations of the synthetic example model's discontinuities with respect to its three independent variables: the algorithm successfully detects the zeroth- and first-order discontinuities in $x_1$ and $x_2$, respectively, but not the second-order discontinuity in $x_3$. (Note that the density is smoothed using a general additive model.)}
    \label{fig:toy-taubar}
\end{figure}

Table~\ref{tab:toyglobal} shows that the global sensitivities when $N = M = 1000$ exhibit a diversity of sensitive and insensitive first-order and total sensitivity indices, thus confirming the requisite behavior desirable for a synthetic example model. The cumulative sensitivities $T_{i,j}^{(2,\epsilon)}(x)$ in Figure~\ref{fig:toy-tx} approach the global sensitivities $\hat{T}_{i,j}$ of Table~\ref{tab:toyglobal} as $x_i \rightarrow 1$. One can also see mild nonlinearities near the $x_1 \doteq 0.59$, and $x_3 \doteq 0.10$ discontinuities of the model:\ this hints at the practicality of decomposing the global sensitivity into local contributions. The local sensitivity density in Figure~\ref{fig:toy-tau} demonstrates the viability of the method and highlights the stronger contributions to the local variance near the model's zeroth- and first-order discontinuities ($x_1 \doteq 0.59$ and $x_2 \doteq 0.95$). Averaging over the output dimensions yields the average local sensitivity density in Figure~\ref{fig:toy-taubar}.

\begin{table}
\centering
\begin{tabular}{l|c|c|c}
$\hat{S}_{i,j}$ & $x_1$ & $x_2$ & $x_3$ \\
\hline
$y_1$ & 0.01 & 0.00 & 0.98 \\
$y_2$ & 0.62 & 0.00 & 0.06 \\
$y_3$ & 0.13 & 0.00 & 0.59 \\
\end{tabular}
\quad
\begin{tabular}{l|c|c|c}
$\hat{T}_{i,j}$ & $x_1$ & $x_2$ & $x_3$ \\
\hline
$y_1$ & 0.01 & 0.00 & 0.99 \\
$y_2$ & 0.94 & 0.00 & 0.38 \\
$y_3$ & 0.41 & 0.00 & 0.87 \\
\end{tabular}
\caption{Global first-order sensitivity (left) and total sensitivities (right) results for the synthetic example model, on a scale from zero (insensitive) to one (sensitive), with one large batch of observations $N = M = 1000$.}
\label{tab:toyglobal}
\end{table}

\begin{figure}
    \centering
    \includegraphics[width=0.300\linewidth]{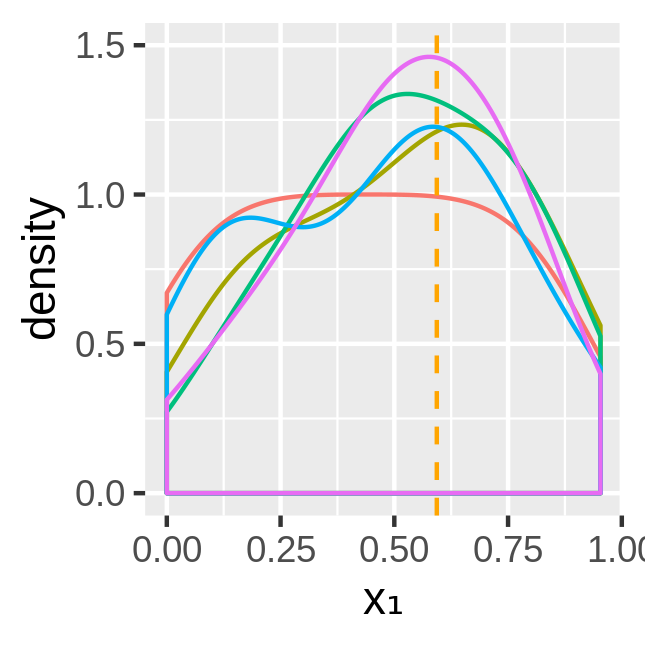}
    \includegraphics[width=0.300\linewidth]{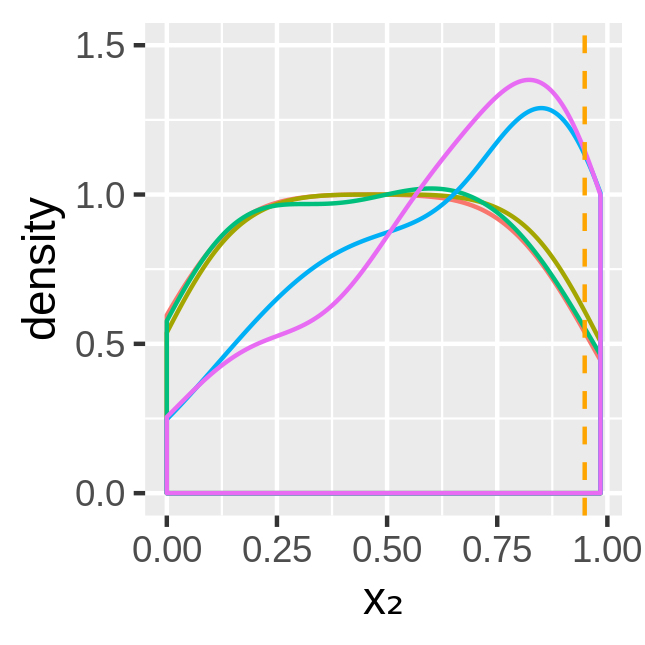}
    \includegraphics[width=0.300\linewidth]{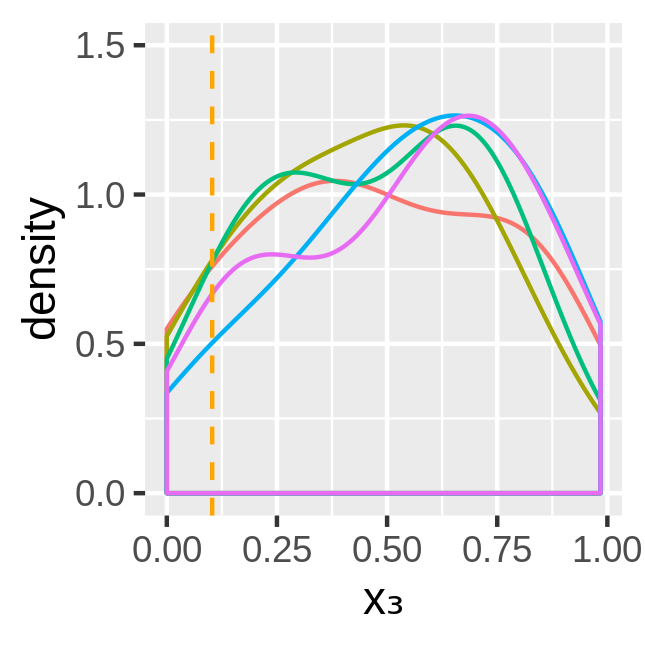}
    \includegraphics[width=0.075\linewidth,clip=true,trim={1.5in 0 2.15 2.15}]{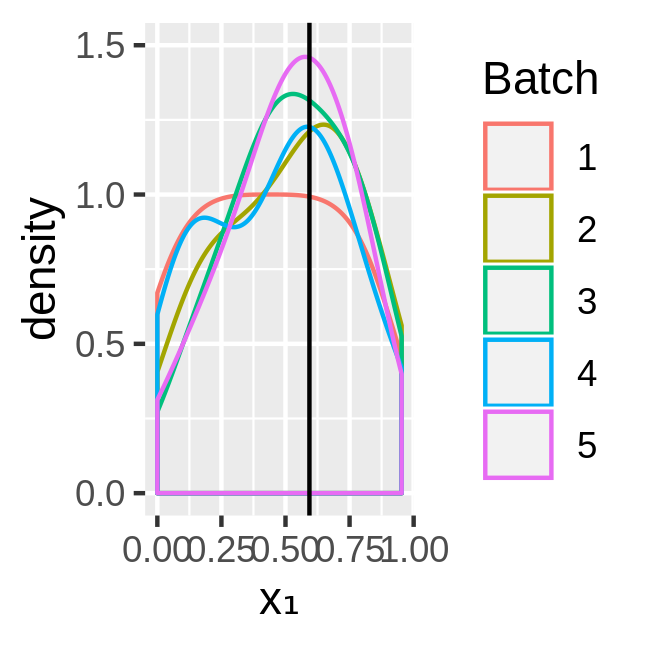}
    \caption{Density of sampling in the first five batches of $M = 10$ observations in adaptive sampling according to $\overline{\tau}_i^{(2,\epsilon)}(x)$ for the synthetic example model, with $\epsilon = 10^{-4}$, starting with a first batch that uniformly samples. The three dashed orange lines mark locations of the synthetic example model's discontinuities with respect to its three independent variables: the algorithm successfully detects the zeroth- and first-order discontinuities in $x_1$ and $x_2$, respectively, but not the second-order discontinuity in $x_3$. (Note that the density is smoothed using a general additive model.)}
    \label{fig:toy-density}
\end{figure}

\begin{figure}
    \centering
    \includegraphics[width=0.300\linewidth]{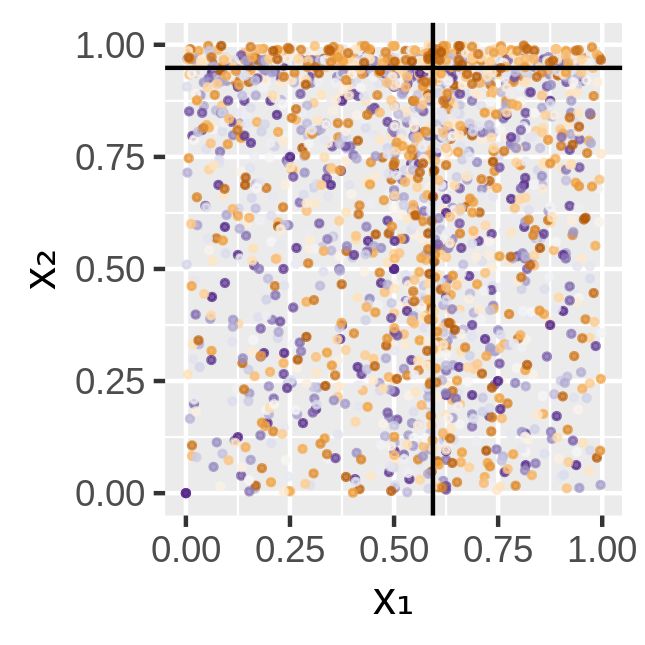}
    \includegraphics[width=0.300\linewidth]{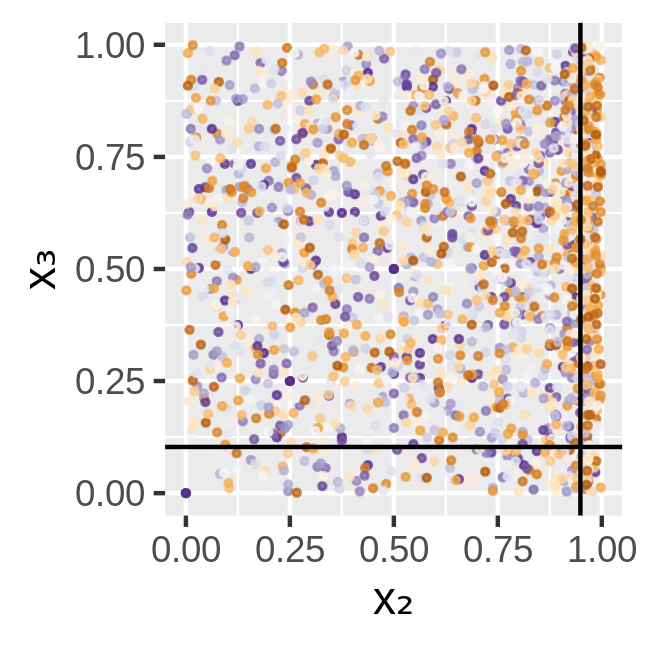}
    \includegraphics[width=0.300\linewidth]{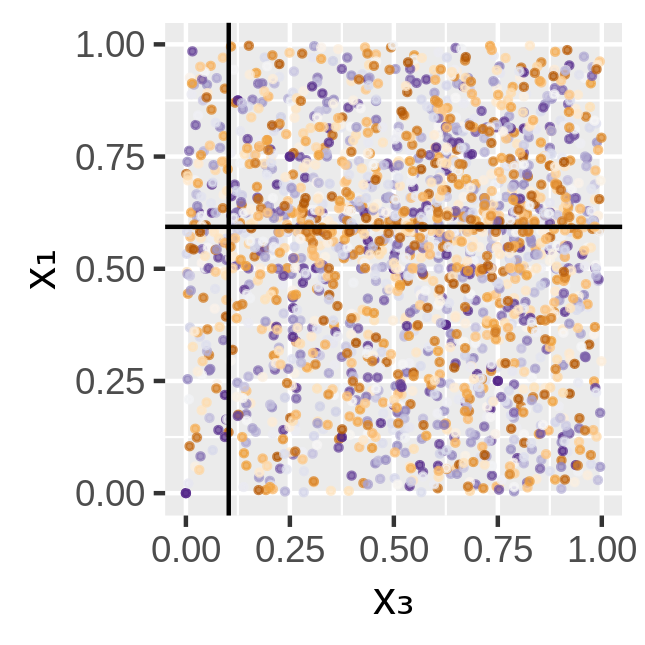}
    \includegraphics[width=0.075\linewidth,clip=true,trim={1.5in 0 2.15 2.15}]{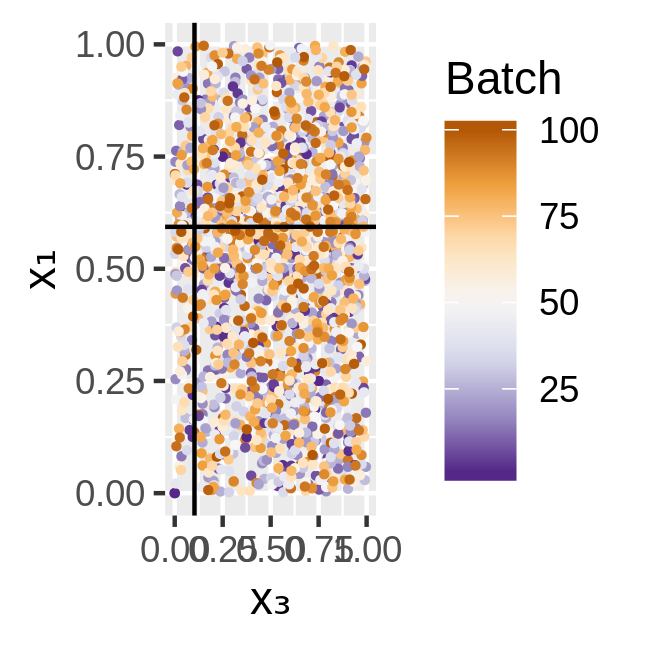}
    \caption{Two-dimensional projections of adaptive sampling according to $\overline{\tau}_i^{(2,\epsilon)}(x)$ for the synthetic example model, with $\epsilon = 10^{-4}$ and for 100 sequential batches of $M = 10$ observations, so $N = 1000$. The solid black lines mark locations of the synthetic example model's discontinuities with respect to its three independent variables: the algorithm successfully detects the zeroth- and first-order discontinuities in $x_1$ and $x_2$, respectively, but not the second-order discontinuity in $x_3$.}
    \label{fig:toy-sample}
\end{figure}

Figure~\ref{fig:toy-density} shows how densely different regions are sampled in the first five batches of $M = 10$ observations when the sensitivity analysis proceeds adaptively. This illustrates how quickly the algorithm detects the areas of higher variance as it transitions from the initial uniform sampling to the sensitivity-emphasizing sampling:\ the first batch samples only slightly non-uniformly, simply because of the few points used from the Sobol' sequence, but by the fifth batch the density forms a noticeable bias towards the higher-variance regions. Moreover, Figure~\ref{fig:toy-sample} displays two-dimensional projections of actual sampling for a batch size of $M = 10$, with 100 batches so $N = 1000$:\ one can see that the algorithm explores the whole space while evolving to concentrate on areas of high sensitivity. Overall, there is no evidence that the algorithm prematurely ``locks into'' local regions of high variance or neglects global exploration of the input space. Employing a higher value of the $\alpha$ exponent in Eq.~(\ref{eq:taux}) would focus sampling near the high-variance regions, while reducing that value to nearly zero would make the sampling more uniform.

\subsection{Bioproduct Transition Dynamics Model}
\label{sec:btd}

We now turn to the demonstration of the adaptive-sampling algorithm on a large, complex bioenergy supply-chain simulation, the Bioproduct Transition Dynamics (BTD) model \citep{hanes_supporting_2020,hanes_transition_2019,hanes_introduction_2018,hanes_system_2018}.
%\todo{\small Should we cite the two papers/reports in preparation?}
The model constitutes a decision support tool for bioproduct industry stakeholders to inform decisions around R\&D investments in and government support of bioproducts. It is a system-dynamics (SD) model consisting of coupled ordinary stochastic differential equations with initial-value boundary conditions.
%\todo{We can count the number of stocks and delays in the BTD to determine how many ODEs it has.}
For the purposes of this example, we consider the subset of the most interesting $k = 84$ input parameters (independent variables) and $n = 49$ output metrics (dependent variables evaluated at the final time of the simulation). We omit further discussion of the model here and just treat it as a ``black box'': see Hanes et al.\ for details of the model (\citeyear{btd_tech_report}) and for a global sensitivity analysis of it (\citeyear{btd_analysis_paper}). Because of the high dimensionality of the input and output spaces, we simply focus here are on few illustrative computations.

\begin{figure}
    \centering
    \includegraphics[width=\linewidth]{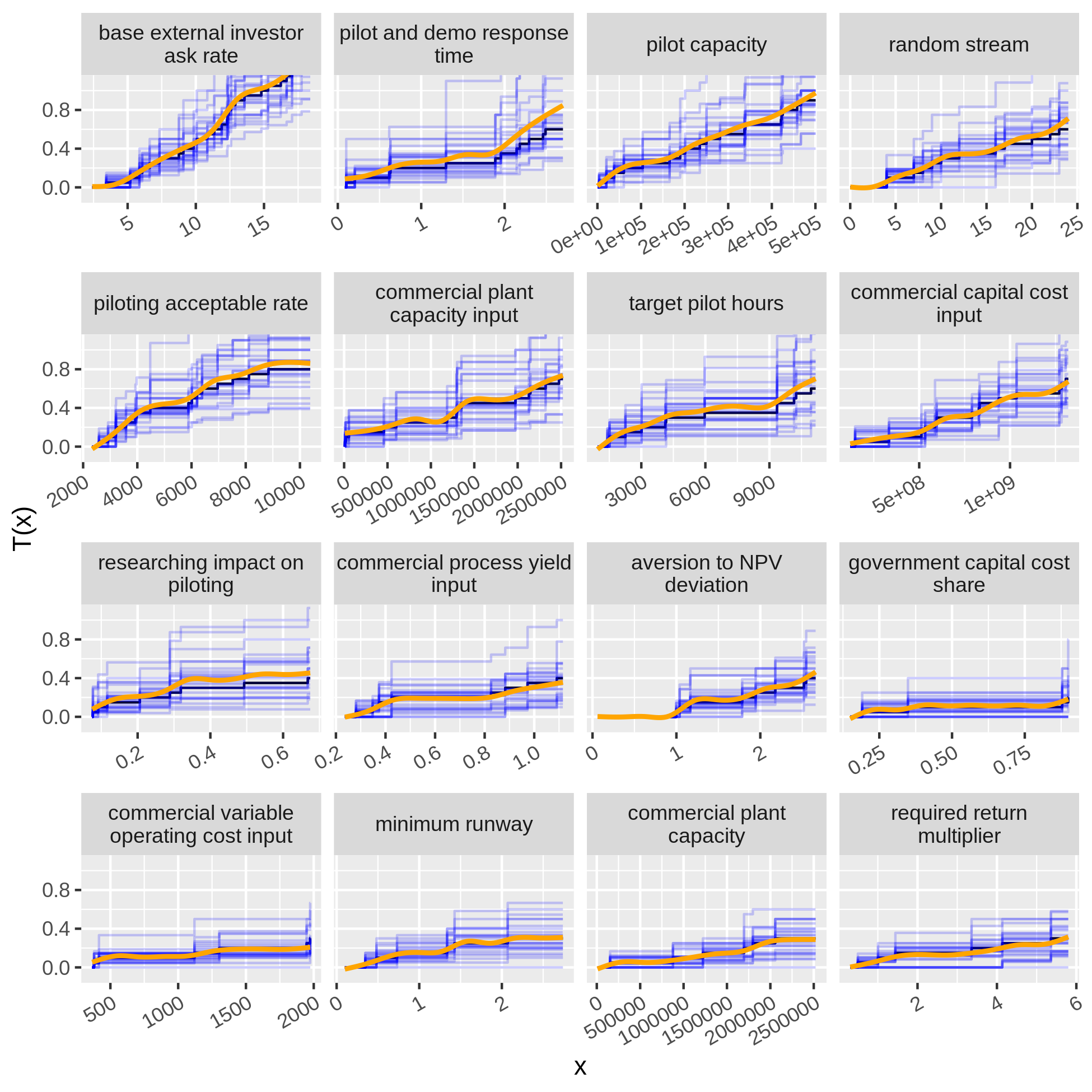}
    \caption{Cumulative local sensitivity $T_{i,j}^{(2,\epsilon)}(x)$ for the BTD model, with $j =$ ``demo plant construction'' and $N = M = 2500$. The black curve is the computed sensitivity, the 25 blue curves are sensitivities obtained by resampling the observations, and the orange curve is the fit from a general additive model for the blue curves.}
    \label{fig:btd-t}
\end{figure}

Figure~\ref{fig:btd-t} plots the cumulative local sensitivity $T_{i,j}^{(2,\epsilon)}(x)$ for the output metric ($j$) ``demo plant construction'' for sixteen representative input parameters ($i$), with $N = 2500$. We see a variety of trends, ranging from flat (non-sensitive), to uniformly increasing (equal contributions to output variance over the input range), to sporadic (concentration of local contributions to the variance). (Here, for the purpose of display, we have selected the strongest sensitivities from the 84 input parameters.) The scatter in the bootstrapped cumulative local sensitivity emphasizes the need for large enough ensemble size in the experimental design. Note, particularly, the sensitivity with respect to the ``random seed'' variable, which represents the initial seed value of the pseudo-random number generator for the BTD's stochasticity:\ one sees the expected linear trend in the cumulative sensitivity, which indicates uniform contributions to the variance as the random seed is changed.

\begin{figure}
    \centering
    \includegraphics[width=\linewidth]{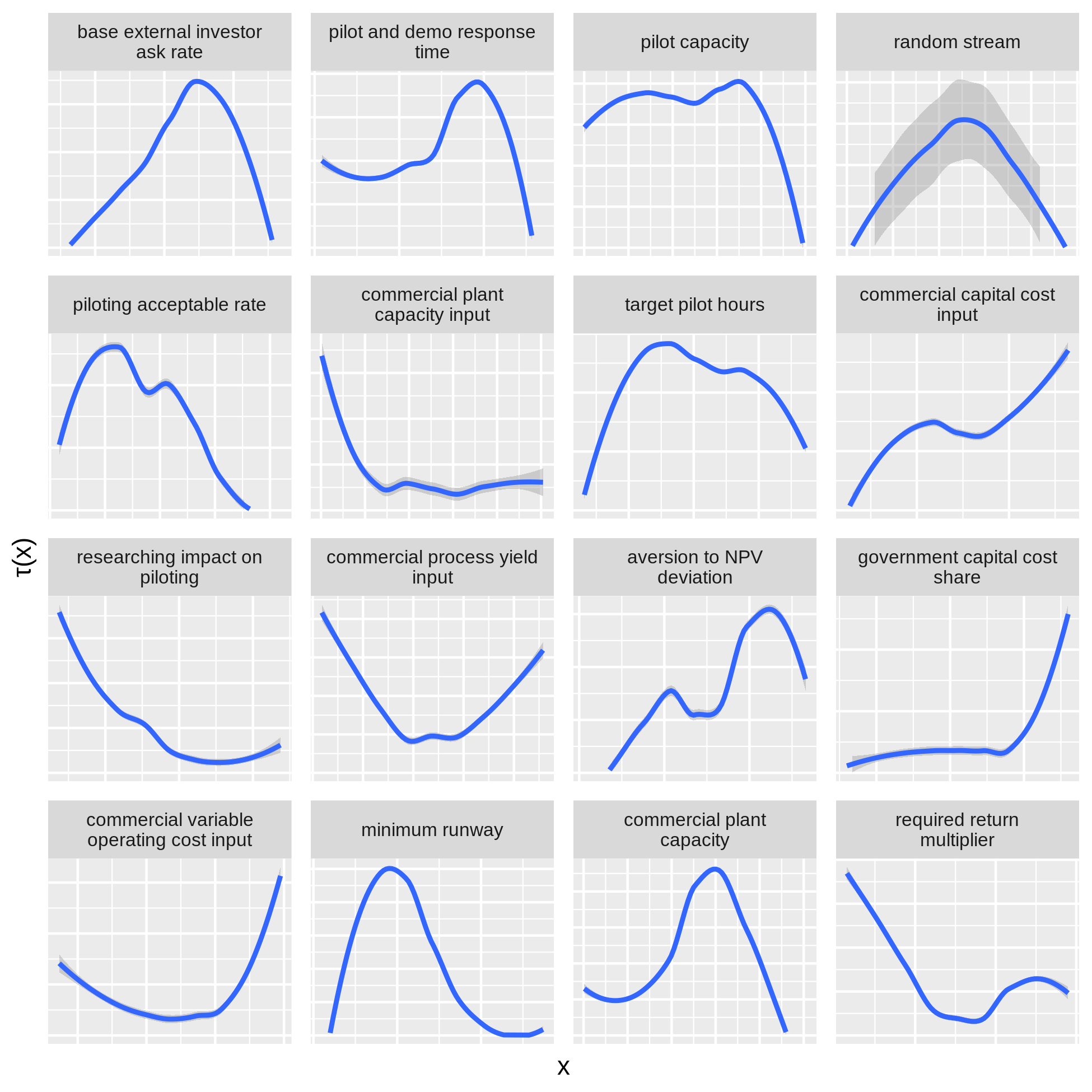}
    \caption{Sensitivity density $\tau_{i,j}^{(2,\epsilon)}(x)$ for the BTD model, with $j =$ ``demo plant construction'' and $N = M = 2500$. (Note that the density is smoothed using LOWESS, with dark gray bands for the confidence interval of the smoothed curve.)}
    \label{fig:btd-tau}
\end{figure}

\begin{figure}
    \centering
    \includegraphics[width=\linewidth]{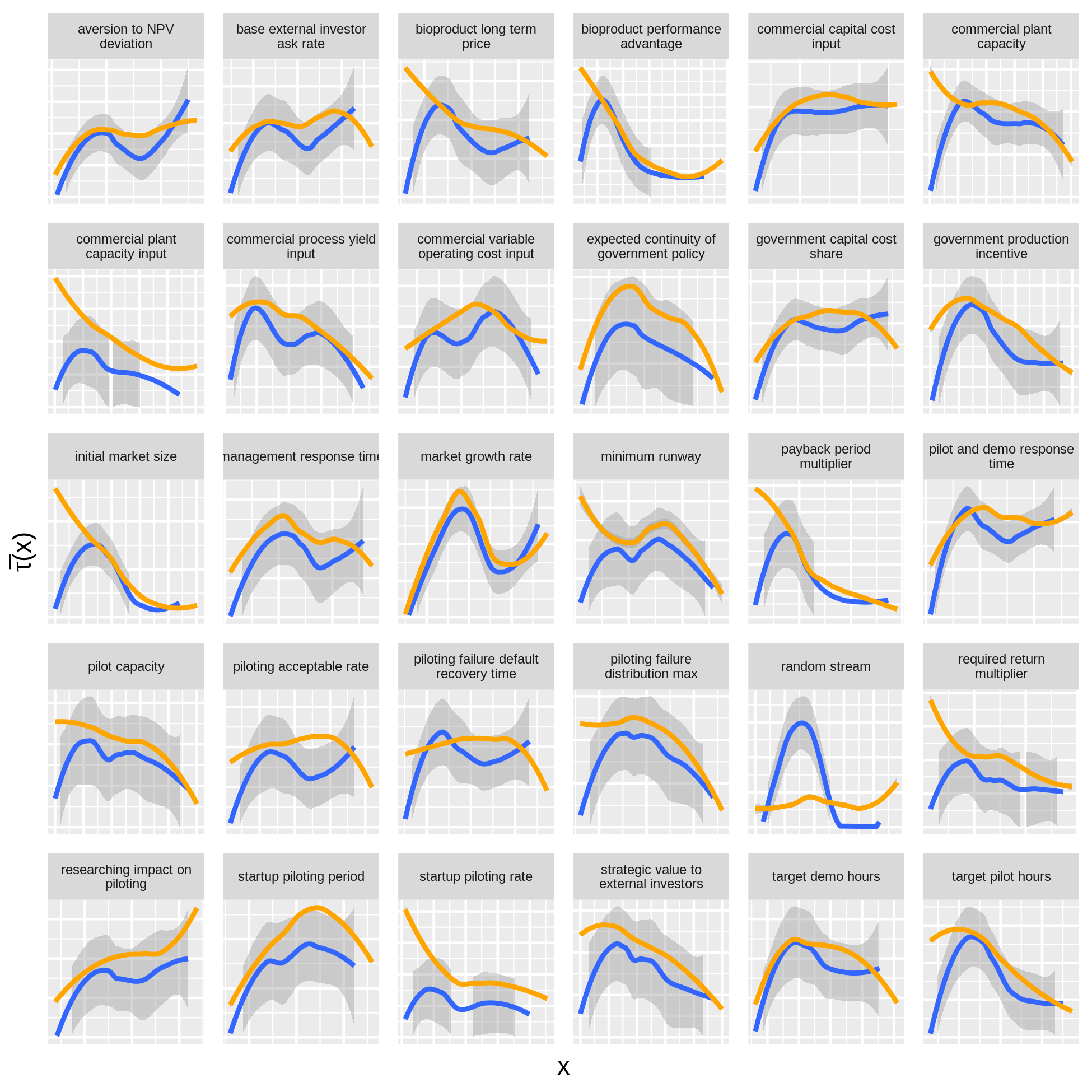}
    \caption{Average sensitivity density $\overline{\tau}_i^{(2,\epsilon)}(x)$ for the BTD model, for the first batch of $M = 10$ observations (blue with dark gray) and for the full set of $N = 2500$ observations (orange). (Note that the density is smoothed using LOWESS, with dark gray bands for the confidence interval of the smoothed curve.)}
    \label{fig:btd-taubar1}
\end{figure}

Transforming the $T_{i,j}^{(2,\epsilon)}(x)$ of Figure~\ref{fig:btd-t} into sensitivity densities, $\tau_{i,j}^{(2,\epsilon)}(x)$, yields the non-uniform sampling in Figure~\ref{fig:btd-tau}. The modes in these densities echo the corresponding high-slope regions in the cumulative sensitivity. When the sensitivity densities for all of the output variables are averaged, Figure~\ref{fig:btd-taubar1} results. (Here, for compactness of display, the 30 strongest sensitivities of the 84 are plotted.) Even the small first batch with $M = 10$ weakly identifies many of the patterns present in the larger $N = 2500$ sample. Comparison with Fig.~\ref{fig:btd-tau} demonstrates that the averaging sometimes mutes the single-output sensitivity density for the ``demo plant construction'' metrics. Notably, the ``random stream'' sensitivity density becomes nearly uniform after averaging when $N = 2500$. As previously noted, altering $\alpha$ of Eq.~(\ref{eq:taux}) would tune the emphasis on high-variance regions.

\section{Discussion}
\label{sec:discuss}

The examples of the synthetic example model and the BTD simulation demonstrate the computational practicality of the method and its ability to identify local input regions of relatively high variance in model output. These examples do not indicate any tendency for the algorithm to lock into a local region of high variance, ignoring the lower variance regions, but they do show that the algorithm detects the higher variance regions in early batches even when the batch size is as low as $M = 10$.

Furthermore, the local sensitivity computations do not rely on partitioning the input, which would result in regions with insufficient numbers of observations within small partitions (either on the boundary of or in the interior of the domain of the independent variable):\ the block/column-replacement experimental design strategy of Sec.~\ref{sec:design} and the interval averaging of Eq.~(\ref{eq:tx}) ensure sufficient observations attributed to each $x$ to keep confidence intervals for the local sensitivity densities generally uniform over $x$. (A contrary formulation of local sensitivity that instead relied on partitioning observations by their independent variable would suffer from sample-size issues in areas where partitions only contain a few observations.)

The method does not explicitly handle time series:\ in the examples, we just use the final time step of the model for the output variables. More sophisticated methods like tensor decomposition \citep{bugbee_enabling_2019} or functional principal components analysis \citep{ramsay_functional_2013} would appropriately handle the multivariate time series that comprise the output in simulation models, but the local-sensitivity method would remain the same after such preprocessing was applied.
%\todo{\small We can include a discussion or parts of Joanne's write-up on function PCA here.}

One substantial deficiency of the method is that it projects $\overline{\tau}_i^{(\alpha,\epsilon)}(x)$ onto the $x_i$ coordinate axes for the inputs. This may make it difficult for the algorithm to detect high-variance regions defined by combinations of inputs supported over a tight interval, especially when the volume of such regions is minuscule compared to the whole input space. Such phenomena would challenge any algorithm, but we anticipate that the fact that Eq.~(\ref{eq:taubar}) is constructed for \textit{total} sensitivity instead of \text{first-order} sensitivity provides an avenue for detecting such small-volume regions once they have been adequately sampled.

Finally, we note that an implementer of the algorithm has the option to adjust $\alpha$ at each batch in the iterative experiment, thus having the ability to tune the intensity of sampling around the high-variance regions. Beyond that, all or part of the density $\overline{\tau}_i^{(\alpha,\epsilon)}(x)$ may be modified or overridden at any iteration according to an analyst's interests in particular input regions. (In fact, densities could also be altered according to non-variance criteria such as in sequential design strategies \citep{box_statistics_2005}.) This all suggests that the approach could be embedded in a larger Bayesian framework that starts with a prior for the sampling density and then updates it both according to the algorithm and to user-specified preferences.

\section{Conclusion}
\label{sec:conclude}

We have constructed a local total sensitivity index that decomposes the global total sensitivity conditions by independent variables, and then shown how this local sensitivity index can be employed in a new method of experimental design that sequentially and adaptively samples the domain of a multivariate function according to local contributions to the global variance. The intensity of sampling of high-variance regions can easily be tuned during the algorithm's iterations. The example application demonstrates performance on a synthetic example model with known discontinuities in its output and to a production-quality simulation used in energy analysis highlighting the practicality and insightfulness of the algorithm.

\if0\blind {
\section*{Acknowledgements}

This work was authored in part by the National Renewable Energy Laboratory, operated by Alliance for Sustainable Energy, LLC, for the U.S. Department of Energy (DOE) under Contract No. DE-AC36-08GO28308. Funding provided by the Department of Energy Office of Energy Efficiency and Renewable Energy Bioenergy Technologies Office. The views expressed in the article do not necessarily represent the views of the DOE or the U.S.\ Government. The U.S.\ Government retains and the publisher, by accepting the article for publication, acknowledges that the U.S.\ Government retains a nonexclusive, paid-up, irrevocable, worldwide license to publish or reproduce the published form of this work, or allow others to do so, for U.S. Government purposes.

Los Alamos National Laboratory, an affirmative action/equal opportunity employer, is operated by Triad National Security, LLC for the National Nuclear Security Administration of U.S. Department of Energy under contract 89233218CNA000001. By approving this article, the publisher recognizes that the U.S. Government retains royalty-free license to publish or reproduce the published form of this article as work performed under the auspices of the U.S. Department of Energy. Los Alamos National Laboratory strongly supports academic freedom and a researcher’s right to publish; as an institution, however, the Laboratory does not endorse the viewpoint of a publication or guarantee its technical correctness.

} \fi

%%%% FIXME: Comment out the following command before submitting.
%\nocite{*}

%%%%\printbibliography
%%%%%\todo{Clean up .bib file.}

\bibliographystyle{jasa3}

\bibliography{sequential-sensitivity}

\appendix

\section{Synthetic Example Model}
\label{sec:toyapp}

This synthetic example model consists of a nonlinear ordinary differential equation for $y_j$ as a function of $x_i$, with $i \in \{ 1 \ldots m \}$ and $j \in \{ 1 \ldots n \}$:
\begin{equation*}
    \frac{dy}{dt} = \kappa \cdot y \ z \ \left( 1 - \sigma \cdot y \  z \right) ,
\end{equation*}
where
\begin{equation*}
    z_j = \sum_{i=1}^m \mathbbm{1}_{j \in L_i} \mathbbm{1}_{x_i \ge \xi_i} \zeta_i \left( x_i - \xi_i \right)^{\delta_i} .
\end{equation*}
The parameters $\xi_i$ define locations of discontinuities and the parameters $\delta_i$ specify the degree of those discontinuities. The parameters $L_i$ affect the mixing of the input variables into the output, as do the matrices $\kappa$ and $\sigma$. The model was designed to exhibit nontrivial and nonlinear behavior with tunable discontinuous behavior. It can be used as a time series generator or just evaluated at a specified final time.

The particular instantiation of the synthetic example model studied in this paper uses the following parameter values:
\begin{align*}
    m & = 3
    \\
    n & = 3
    \\
    t & \in [0, 10]
    \\
    L & = \begin{bmatrix} \left\{ 2, 3 \right\} & \left\{ 1, 2 \right\} & \left\{ 1, 2, 3 \right\} \end{bmatrix}
    \\
    \delta & = \begin{bmatrix} 0 & 1 & 2 \end{bmatrix}
    \\
    \xi & = \begin{bmatrix} 0.5933 & 0.9485 & 0.1030 \end{bmatrix}
    \\
    \zeta & = \begin{bmatrix} 0.8788 & 0.2668 & 0.6661 \end{bmatrix}
    \\
    y(0) & = \begin{bmatrix} -0.1900 & 0.5145 & 0.4094 \end{bmatrix}
    \\
    \kappa & = \begin{bmatrix} 0.7054 & 0.2921 & 0.7361 \\ -0.1151 & 0.5206 & -0.0707 \\ 0.3475 & -0.0579 & -0.2229 \end{bmatrix}
    \\
    \sigma & = \begin{bmatrix} 0.0294 & 0.1668 & 0.5788 \\ 0.1046 & 0.1705 & 0.2749 \\ -0.1258 & -0.0712 & 0.7372 \end{bmatrix}
\end{align*}

\if0\blind {
The following R code, developed by the authors at the National Renewable Energy Laboratory, implements this synthetic example model:
} \fi
\if1\blind {
The following R code implements this synthetic example model:
} \fi

\begin{small}
\begin{verbatim}
# Create a multivariate function with specified properties:
#   tmax: maximum time
#   multiplicities: number of correlations each parameter has
#   degrees: polynomial degree of each parameter
#   dimension: the dimension of the output
#   returns a multivariate function of the vector of parameters and time
makeGenerator <- function(multiplicities, degrees, dimension) {
    single <- function(degree) {
      x0 <- runif(1)
      z0 <- runif(1)
      print(paste("Critical point at x = ", x0, sep = ""))
      function(x) {
          if (x < x0)
              0
          else
              z0 * (x - x0)^degree
      }
    }
    locations <- lapply(
        multiplicities,
        function(m) sample(1:dimension, m)
    )
    functions <- lapply(degrees, single)
    start <- runif(dimension, -0.25, 0.75)
    coefs <- matrix(
        runif(dimension^2, -0.25, 0.75),
        dimension,
        dimension
    )    
    shift <- matrix(
        runif(dimension^2, -0.25, 0.75),
        dimension,
        dimension
    )
    function(x, ts) {
        z <- rep(0, dimension)
        for (i in 1:length(locations))
            for (j in locations[[i]])
                z[j] <- z[j] + functions[[i]](x[i])
        ode(start, ts, function(t, y, params) {list((coefs %*% y) * z *
            (1 - ((shift %*% y) * z)))})
    }
}
\end{verbatim}
\end{small}

Here is a simple example of the use of this function:
\begin{small}
\begin{verbatim}
# Use reproducible random numbers.
RNGkind("Mersenne-Twister", "Inversion", "Rejection")
set.seed(46)

# Instantiate the model.
f <- makeGenerator(c(2, 2, 3), c(0, 1, 2), 3)
#   "Critical point at x = 0.593385165324435"
#   "Critical point at x = 0.948547213338315"
#   "Critical point at x = 0.102978735696524"

# Evaluate it at x = (0.1, 0.2, 0.3) for t = 0, 5, 10.
f(c(0.1, 0.2, 0.3), c(0, 5, 10))
# time    1          2          3
# 0     -0.1900320  0.5144967  0.4093612
# 5     -0.1478757  0.5489932  0.3864914
#10     -0.1024813  0.5854096  0.3659173
\end{verbatim}
\end{small}

\end{document}